\documentclass[pdflatex, sn-nature, iicol]{sn-jnl}% Math and Physical Sciences Numbered Reference Style
\setcitestyle{super}

\usepackage{graphicx}%
\usepackage{multirow}%
\usepackage{amsmath,amssymb,amsfonts}%
\usepackage{amsthm}%
\usepackage{mathrsfs}%
\usepackage[title]{appendix}%
\usepackage{xcolor}%
\usepackage{textcomp}%
\usepackage{manyfoot}%
\usepackage{booktabs}%
\usepackage{algorithm}%
\usepackage{algorithmicx}%
\usepackage{algpseudocode}%
\usepackage{listings}%

\usepackage[T1]{fontenc}
\usepackage{comment}
\usepackage{siunitx}
\usepackage{hyperref}
\usepackage{float}

\usepackage{pdfpages}
\begin{document}

\title[Article Title]{Absolute Primary Nanothermometry Using Individual Stark Sublevels of Rare-Earth-doped Crystals}

\author*[1,2]{\fnm{Allison R.} \sur{Pessoa}}\email{allison.pessoa@ufpe.br}

\author[3]{\fnm{Thomas} \sur{Possmayer}}

\author[1]{\fnm{Jefferson A. O.} \sur{Galindo}}

\author[4]{\fnm{Luiz F. dos} \sur{Santos}}

\author[4]{\fnm{Rogéria R.} \sur{Gonçalves}}

\author[3,1]{\fnm{Leonardo de S.} \sur{Menezes}}

\author[1]{\fnm{Anderson M.} \sur{Amaral}}%\email{anderson.amaral@ufpe.br}

\affil[1]{\orgdiv{Department of Physics}, \orgname{Universidade Federal de Pernambuco}, \orgaddress{\city{Recife}, \postcode{50740-540}, \state{Pernambuco}, \country{Brazil}}}

\affil[2]{\orgname{Instituto Federal de Educação, Ciência e Tecnologia de Pernambuco}, \orgaddress{\city{Recife}, \postcode{50740-545}, \state{Pernambuco}, \country{Brazil}}}

\affil[3]{\orgdiv{Chair in Hybrid Nanosystems, Faculty of Physics}, \orgname{Ludwig-Maximilians-Universität München}, \orgaddress{\city{Munich}, \postcode{80539}, \state{Bavaria}, \country{Germany}}}

\affil[4]{\orgdiv{Department of Chemistry, Faculdade de Filosofia, Ciências e Letras de Ribeirão Preto}, \orgname{Universidade de São Paulo}, \orgaddress{\city{Ribeirão Preto}, \postcode{14040-900}, \state{São Paulo}, \country{Brazil}}}

\abstract{We present two independent optical methods for absolute primary thermometry using rare-earth-doped nanoparticles. Both approaches rely exclusively on the internal energy levels and population dynamics of the dopant ions, eliminating the need for external temperature references. We experimentally demonstrate the concepts by using Y$_2$O$_3$: Yb$^{3+}$/Er$^{3+}$ nanoparticles, exploiting Boltzmann distribution between individual Stark sublevels of the Er$^{3+}$ ions, emitting in the green spectral region ($\sim$550 nm) and in the near-infrared spectral region ($\sim$1600 nm). Our strategy establishes rare-earth-based luminescence thermometers as genuine absolute primary probes, conceptually comparable to Johnson noise and acoustic gas thermometers, but with the fundamental advantage of possibly being employed at the nanoscale, potentially down to the single-ion limit, with optical readout and over wide temperature ranges.}


\maketitle

\section{Introduction}

Temperature is a fundamental physical quantity that governs phenomena such as chemical reactivity, phase transitions, and biological processes. Thermometers that rely on a well-defined physical law (equation of state) relating the thermometric parameter ($R$) to the thermodynamic temperature ($T$) are termed as \textit{primary thermometers} \cite{BIPM_2019}. In this case, the relationship between $T$ and $R$ is ideally independent of unknown temperature-dependent variables. With the Kelvin scale redefinition in 2019 \cite{BIPM_2019}, the Bureau International des Poids et Mesures (BIPM) introduced the terminology `absolute primary thermometry' for measurement techniques that determine $T$ without reference to any external temperature, and `relative primary thermometry' for techniques that require one or more fixed temperature points to establish key parameters of the equation of state. Standard examples of absolute primary thermometers include acoustic gas thermometers (AGT) \cite{Moldover_2014}, Johnson noise thermometers (JNT) \cite{Benz_2024} and blackbody radiation-based thermometers \cite{Hartmann_2011}. In particular, AGTs are currently used for the state-of-the-art determination of the triple point of water, achieving relative uncertainties on the order of 10$^{-6}$ \cite{Moldover_2014}. More recently, Brillouin scattering by gases in hollow-core optical fibers has been also proposed for absolute primary approaches \cite{Yang_2026}. 

At the nanoscale, however, accurate temperature measurements pose a significant challenge due to the limited spatial resolution of conventional thermometers and the invasive nature of most contact-based techniques \cite{Brites_2012}. As a result, there has been a growing interest in the development of luminescent nanothermometers that can convert temperature changes into optical signals in a minimally invasive fashion \cite{Dacanin_2023}. Among various luminescent nanothermometric platforms, such as quantum dots \cite{Haupt_2014}, organic dyes \cite{Kalaparthi_2021} and single nanodiamonds containing color centers\cite{Neumann_2013}, rare-earth-doped nanoparticles have emerged as particularly attractive candidates \cite{Bradac_2020}. These systems benefit from the unique optical properties of trivalent lanthanide ions (Ln$^{3+}$), including sharp emission lines, long excited-state lifetimes, and high photostability \cite{Zhang_2022, Dong_2020}. Thermometry using Ln$^{3+}$-doped particles has been demonstrated across a wide temperature range, spanning from cryogenic temperatures ($>$ 2.5 K) \cite{Boldyrev_2024_LiYF4, Boldyrev_2024, Possmayer_2026, Kulesza_2025, Dechao_2021} to temperatures exceeding 1000 K \cite{Kulesza_2025, Runowski_2020, Hernandez_2021, Dechao_2021}. In particular, Yb$^{3+}$/Er$^{3+}$-codoped crystalline solids are widely used for relative primary thermometry around room temperature \cite{Suta_2025}, for which most studies rely on the Er$^{3+}$ thermally coupled spin–orbit manifolds $^2$H$_{11/2}$ and $^4$S$_{3/2}$, emitting in the green spectral region \cite{Goncalves_2021}. More recently, Possmayer \textit{et al.} \cite{Possmayer_2026} exploited the Stark sublevel structure of the $^4$S$_{3/2}$ manifold for relative primary thermometry in the cryogenic range. Similarly, Boldyrev \textit{et al.} \cite{Boldyrev_2024} investigated the Stark structure of the $^4$I$_{13/2}$ manifold emitting in the near-infrared range (NIR) also for cryothermometry.

Despite their advantages, most current implementations of Ln$^{3+}$-based luminescent nanothermometers are limited to relative primary methods \cite{Balabhadra_2017, Goncalves_2021, Martins_2023}. In practice, this means that the thermometric parameter must be calibrated against one or more temperature points, determined by external probes. This approach assumes that the calibration remains valid under possibly varying experimental conditions. However, several factors are known to influence the calibration parameters, both in ensemble experiments \cite{Alencar_2004} and, in particular, at the single-nanoparticle level. The latter includes effects of the surrounding medium \cite{Galindo_2021, Galindo_2021_corr}, photonic effects in the host medium \cite{Vonk_2023, Stopikowska_2023}, and the presence of power-dependent intruding luminescence bands \cite{Pessoa_2023, Ruhl_2021}, all of which may compromise the reliability of such approaches.

Towards the realization of absolute primary thermometers working at the nanoscale and over a wide temperature range, the work of Souza \textit{et al.}\cite{Souza_2016} represents an important step using Eu$^{3+}$-doped systems. The authors exploit the Boltzmann population distribution among closely spaced spin-orbit levels near the Eu$^{3+}$ ground state. By tuning the excitation wavelength from 570 nm to 640 nm, they selectively addressed the initial manifold $\langle i|$ involved in the optical transition $\langle i |$ $\rightarrow$ $^5D_0$. The thermodynamic temperature was then determined from the emission spectra of the $^5D_0$ $\rightarrow$ $^7F_4$ transition after the controlled excitation. Because the method depends on this specific arrangement of energy levels, extending it to other Ln$^{3+}$ ions is not straightforward. Here, we propose two alternative procedures for realizing absolute primary optical nanothermometry that rely exclusively on the emission spectrum, requiring no variation in excitation wavelength. We demonstrate both approaches by exploiting the Stark sublevel structure of Er$^{3+}$ ions in Y$_2$O$_3$: Yb$^{3+}$/Er$^{3+}$ nanoparticles under 980 nm illumination, thoroughly analyzing their respective advantages and limitations.

\section{Results}
%\subsection{Manifold thermalization}
\subsection{Proposed methods for absolute primary thermometry using Stark sublevels}

When Ln$^{3+}$ ions are doped into crystalline hosts, their degenerate spin-orbit manifolds are split into closely spaced Stark sublevels, typically separated by energies on the order of \qty{e2}{\per\centi\meter} \cite{Hanninen_2010}. Upon electronic excitation to a given manifold, electrons rapidly thermalize among its Stark sublevels (and possibly with adjacent manifolds) on the time scale of picoseconds to nanoseconds, which is usually much shorter than the characteristics radiative lifetimes \cite{Meltzer_2001, Pessoa_2025}. The thermally accessible levels within a given temperature range are said to be Thermally Coupled (TC). Radiative transitions from the Stark sublevels of the excited manifold to sublevels of lower-lying manifolds give rise to luminescence bands composed of many Stark-Stark spectral lines. Under thermal equilibrium conditions, Boltzmann distribution dictates that the luminescence intensity ratio (LIR) between two TC Stark spectral lines follows a single exponential function \cite{Pessoa_2025}

\begin{equation}
R_\text{Stark}(T) = C\cdot \exp\left(-\frac{\Delta E}{k_\text{B} T}\right)   \quad ,
\label{eq:LIR_Stark}
\end{equation}

\noindent where $C$ is a constant that depends on the radiative decay rates from the two emitting TC sublevels and $\Delta E$ is their energy difference. 

To realize an absolute primary thermometer based on the Boltzmann distribution among the Stark sublevels, it is necessary to know $\Delta E$ and $C$ from Eq. \eqref{eq:LIR_Stark} independently of external thermometers. By analysing the peak position of the lines in the luminescence spectrum, we can confidently determine $\Delta E$ with uncertainties on the order of \qty{0.5}{\per\centi\meter} or less\cite{Possmayer_2026}, limited by the host matrix and/or by the resolution of the spectrometer used for recording the spectra. On the other hand, when measuring the spectrum with photon-counting devices, as with CCD cameras, the parameter $C$ can be written as $C = g_2A_{20}/g_1A_{10}$, where $g_{2}(g_{1})$ is the degeneracy of the higher(lower)-lying TC sublevel and $A_{20}$($A_{10}$) is the radiative transition rate of the corresponding Stark-Stark transition \cite{Pessoa_2025, Suta_Meijerink_2020}. Calculating these rates from first principles is difficult for several reasons. For instance, electric dipole transitions are parity-forbidden between $4f$ orbitals, such that the radiative emission requires a mixing of several energy levels \cite{Walsh_2006, Hovhannesyan_2022}. Also, the local crystal field influences both the energy levels and the electric dipole moments with distinct dependences \cite{Hehlen_2013}. While first principles calculations are challenging for a real sample, in the following sections we develop two alternative, self-referenced, experimental strategies for obtaining the parameter $C$ based on the thermal population distribution of the TC levels.

To develop the self-referenced procedures, we first need to assume that the particles are in direct thermal contact with a heating/cooling element, which heat transfer rate $\dot{Q}$ can be controlled. The rate $\dot{Q}$ can also be represented by an external parameter $P$, for instance, as the position of the control knob for the electrical power dissipated in a resistive heater or in a Peltier module, provided that the relation between $P$ and $\dot{Q}$ is linear. The heat source will produce temperature gradients across the composite system (particles + surroundings), depending on the thermal resistances of the materials. The exact temperature at the particle's location is not necessarily known beforehand, but for suitable temperature ranges it will be proportional to the rate $\dot{Q}$ (see section S1 of the Supplementary Information (SI) material for a detailed analysis). In other words, the temperature of the particle, $T$, varies linearly with the heat transfer rate $\dot{Q}$, consequently also with the control parameter $P$, as 

\begin{equation}
T = a P + T_0   \quad ,
\label{eq:T_linearized}
\end{equation}

\noindent where $a$ indicates the effectiveness of the control knob in tuning the particle's temperature, being determined by the thermal resistance and boundary conditions of the system. $T_0$ corresponds to the equilibrium temperature of the thermometer when the control system is off. When the system's effective thermal resistances change with temperature, a relation as Eq. \eqref{eq:T_linearized} should also be verified around a smaller temperature range of interest. We remark that our methodological proposal depends only on the existence of a linear relation between $T$ and $P$ for a given region used for the self-referenced calibrations, but not on the specific values of the constants $a$ and $T_0$.

In this work, the heating device is a closed-cycle cryostat that allows tuning the electrical power applied to its internal resistive elements. It acts to increase the temperature inside the cryocooler chamber above the liquid helium boiling point (4.2 K). Since in our case we can directly tune the electrical power delivered by the resistive elements, we can make $P \propto \dot{Q}$, therefore associating $P$ to the heat transfer rate. Section S2 of the SI shows the linearity between $P$ and the cryostat internal temperature, measured with an independent thermometer (see the details in the Materials and Methods section). Given the validity of Eq. \eqref{eq:T_linearized}, we discuss in what follows that $C$ can be obtained by studying the relation between $R_\text{Stark}$ and $P$ at sufficiently high temperatures (Subsec. 2.1.1) and also by exploiting the inflection point in the optimal sensitivity region (Subsec. 2.1.2).

\subsubsection{Self-referenced calibration using the high temperature limit}

\begin{figure*}[!ht] % Double column figure
\begin{center}
	\includegraphics[width=\textwidth]{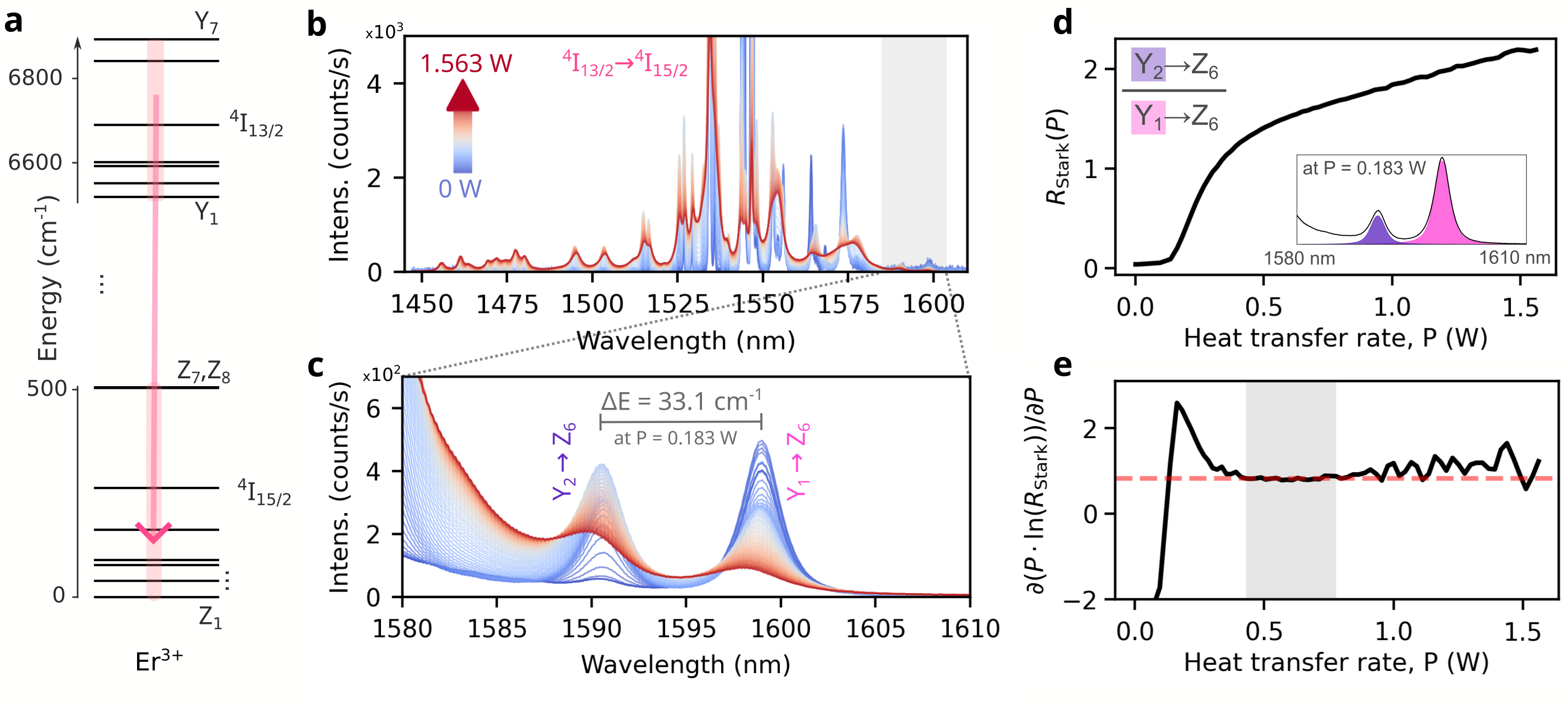}
	\caption{\textbf{a}: Simplified energy-level diagram of Er$^{3+}$ ions in Y$_2$O$_3$ matrix. The arrow and shaded regions represent spontaneous radiative decay from all Stark sublevels of $^4$I$_{13/2}$ to those of $^4$I$_{15/2}$. \textbf{b}: Resulting luminescence spectra of the electronic transitions as a function of the heat transfer rate, $P$. Y-axis was limited to improve reading. \textbf{c}: $P$ sweep of the two particular Stark-Stark lines (Y$_1$ $\rightarrow$ Z$_6$ and Y$_2$ $\rightarrow$ Z$_6$) chosen as thermometric probes. The energy centers of the lines were measured at an arbitrary $P$ point according to the methods described in section S3. \textbf{d}: Measured $R_\text{Stark}$ as a function of $P$, for the full working range of the cooling/heating device (cryostat). Inset shows a spectrum at an arbitrary $P$ point with the Voigt fittings of the Stark lines used to calculate $R_\text{Stark}$. \textbf{e}: Derivative of the experimental curve $P \cdot \ln R_\text{Stark}$, demonstrating linearity in the limit of high temperatures. Shaded region shows the limits used to calculate the average of the points (red dashed line).}
	\label{fig:spectrum_Stark_IR}
\end{center}
\end{figure*}

For thermal energies much higher than the energy difference between the TC sublevels ($k_\text{B}T >> \Delta E$), the occupation probability asymptotically becomes 50\% for both, which can be used to estimate $C$. In this case, $\exp(-\Delta E/ k_\text{B}T) \approx 1$, so $R_\text{Stark} \approx  C = g_2A_{20}/g_1A_{10}$. 

If $C$ or $\Delta E$ do not change significantly at high temperatures, it is possible to experimentally extrapolate the asymptotic limit for $R_\text{Stark}(P)$ from measurements at sufficiently large temperatures, therefore extracting $C$. This is achieved by measuring a set of spectra for various control parameters $P$, obtaining a $R_\text{Stark}(P)$ curve. Although $R_\text{Stark}(P)$ has a complicated dependence with $P$, a better experimental relation would be obtained by taking the product $P \ln R = P \ln C - \Delta E/k_\text{B} \times P/(a P + T_0)$. While this expression seems too complex, a plot of $P \ln R$ \textit{vs.} $P$ is such that the slope at a given point can be written as

\begin{equation}
\frac{\mathrm{d}(P \cdot \ln R_\text{Stark})}{\mathrm{d}P} = \ln C - \frac{\Delta E}{k_\text{B}T}\frac{T_0}{T}   \quad .
\label{eq:derivative_composite}
\end{equation}

\noindent That is, the asymptotic slope of $P \cdot \ln R_\text{Stark}$ in the high-temperature limit can be used to extract $\ln C$. Notice that the curve can be represented using directly the experimental data and we have not referred to the sample's temperature in any step, neither is necessary to determine $a$ or $T_0$. Thus, even though the thermal resistances of the system may change with $T$, our proposal remains valid to extract $C$ as long as $C$ itself does not change significantly with temperature. The linearized parameters $a$ and $T_0$ will change with the calibration mean setpoint temperature, but they do not influence significantly the estimate of $C$ around sufficiently large particle's temperatures $T$. Since the temperature-dependent term in Eq. \eqref{eq:derivative_composite} scales with $T^{-2}$, the linear slope can be attained for relatively low temperatures provided that $\Delta E / k_\text{B} \ll T$ and $T_0 \ll T$. Also notice that the $T^{-2}$ dependence suggests that Eq. \eqref{eq:derivative_composite} converges faster to $\ln C$ instead of using $\ln R_\text{Stark}$ only. 

Performing the measurements at high temperature ranges reduces inaccuracy, but clearly the thermal stability of the host matrix limits the maximum temperature it supports without changing its properties (including possible phase transitions at high temperatures). Therefore, the energy difference between the TC Stark sublevels should be low enough to allow reaching the `high' temperature limit without significant changes in the host matrix. Naturally, this limits the range of optimal thermal response \cite{Suta_Meijerink_2020}, making this self-referenced calibration procedure very efficient to thermometers at cryogenic temperature ranges. From the practical point-of-view, at higher temperatures blackbody radiation might intrude and may need to be properly subtracted. Stark lines will spectrally broaden and might not be as easily separable. Intruding bands can also be thermally activated and thus disrupt the $R_\text{Stark}$ measurement \cite{Pessoa_2023}.

To demonstrate this strategy, we placed the Y$_2$O$_3$: (1.5\%) Yb$^{3+}$ / (0.5\%) Er$^{3+}$ nanoparticles inside a heating/cooling device (cryostat). The sample was excited with with laser illumination at 980 nm, and the luminescence emission in the NIR, around 1550 nm, was collected (see Materials and Methods section). This luminescence band corresponds to transitions between the Stark sublevels of the $^4$I$_{13/2}$ manifold of the Er$^{3+}$ ions to the Stark sublevels of the ground-state manifold ($^4$I$_{15/2}$) \cite{Boldyrev_2024}. In particular, we used the luminescence emission of the two lowest Stark sublevels of $^4$I$_{13/2}$, labeled as Y$_1$ and Y$_2$ \cite{Gruber_2008}. These sublevels are separated by only 33.08 $\pm$ 0.05 cm$^{-1}$ at 20 K (see section S3 of the SI), thus making the optimum temperature measurement window of 14--24 K \cite{Suta_Meijerink_2020}. The simplified energy-level structure of the $^4$I$_{13/2}$ manifold is shown in Fig. \ref{fig:spectrum_Stark_IR}\textbf{a}. The resulting observed luminescence spectra after laser excitation at 980 nm are shown in Fig. \ref{fig:spectrum_Stark_IR}\textbf{b} for heat transfer rates from 0 up to 1.563 W (the maximum the device can provide). This power range corresponds to internal temperature variations between 4.4 K and 350.0 K. 

We selected the Stark pair Y$_1$ $\rightarrow$ Z$_6$ and Y$_2$ $\rightarrow$ Z$_6$, located between 1585 and 1605 nm, because they are well separated from neighboring peaks (Fig. \ref{fig:spectrum_Stark_IR}\textbf{c}). This spectral isolation enables unambiguous transition assignment and reliable Voigt-profile fitting (see Section S3 of the SI for details on the deconvolution procedure). Additionally, this wavelength range lies within the telecommunication L band, and also in the NIR III optical window for biological tissues, which can be important to a variety of applications.

In Fig. \ref{fig:spectrum_Stark_IR}\textbf{d} we show the $R_\text{Stark}(P)$ curve for the full working range allowed by the heating system. Based on Eq. \eqref{eq:derivative_composite}, we can determine $\ln C$ by the slope of the $P\cdot \ln R_\text{Stark}$ \textit{vs.} $P$ curve in the asymptotic limit of $P\rightarrow \infty$. Fig. \ref{fig:spectrum_Stark_IR}\textbf{e} depicts the first derivative of $P\cdot \ln R_\text{Stark}$ with respect to $P$, showing that for $P \rightarrow \infty$, the curve resembles a straight line. To extract $\ln C$, we used an average of 22 data points of Fig. \ref{fig:spectrum_Stark_IR}\textbf{e} between $P = 0.43$ W and $P = 0.78$ W (shaded region) where $R_\text{Stark}$ presents a reasonable signal-to-noise ratio. We therefore obtain $C = 2.3 \pm 0.1$. For higher $P$ values, the overall intensity of the luminescence lines decreases while their width increases, difficulting spectral separation thus increasing uncertainty. We notice that for $P > 0.78$ W, the curve derivative slightly increases, most probably due to the low quality of the spectral separation. The uncertainty was calculated based on conventional error propagation from the fitting (see section S5 of the SI). This value is in agreement with that determined by temperature-dependent verification (Fig. S4\textbf{b}).

\begin{figure}[!h] % SinSignal gle column figure
\begin{center}
	\includegraphics[width=\columnwidth]{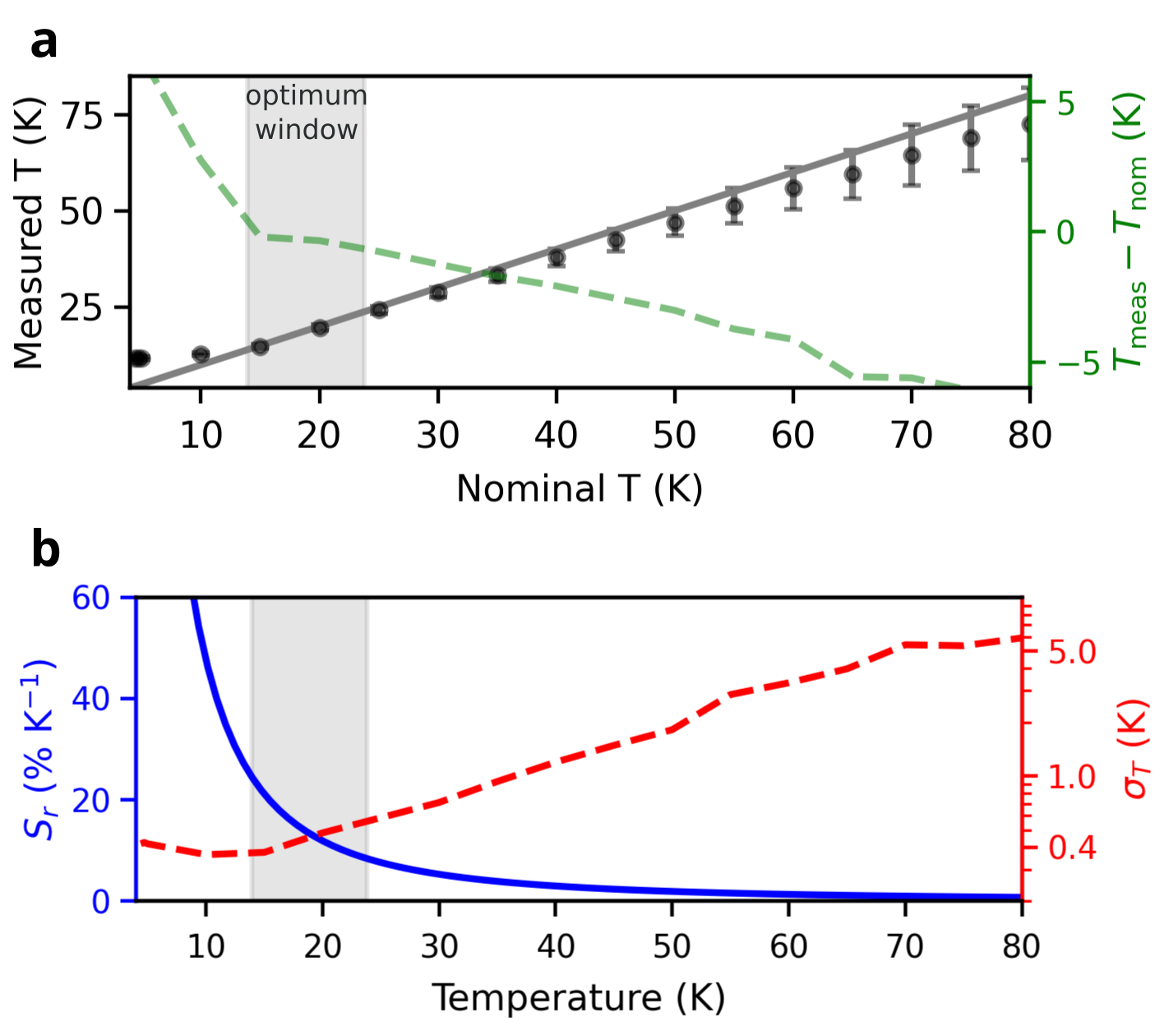}
	\caption{\textbf{a}: Comparison between temperature measurements of the proposed optical thermometer (measured T) and the internal cryostat high-resolution reference (nominal T). The solid straight line represents the function $y=x$. The right-hand y-axis shows the difference between these two quantities. \textbf{b}: Characterization of the optical nanothermometer showing the relative sensitivity ($S_r$) and thermal uncertainty ($\sigma_T$).}
	\label{fig:thermometer_IR_charac}
\end{center}
\end{figure}

By measuring $R_\text{Stark}$ and $\Delta E$ for a given spectrum, and considering $C = 2.3 \pm 0.1$, one has all necessary parameters to obtain $T$ from Eq. \eqref{eq:LIR_Stark}. In Fig. \ref{fig:thermometer_IR_charac}\textbf{a} we show the measured thermodynamic temperatures and the nominal ones, determined with an independent probe (the cryostat internal thermometer) for validation. The right-hand axis shows the difference between the temperature measured using the method just described and the cryostat reference. This represents the accuracy of our thermometer, since the cryostat manufacturer guarantees an accuracy of $\pm$ 10 mK around 30 K \cite{LakeShore_AppendixD_2024}. Our optical thermometer shows a maximum accuracy of $\pm$ 0.5 K for temperatures within its optimum window. In Fig. \ref{fig:thermometer_IR_charac}\textbf{b} we show the relative sensitivity $S_r(T) = R^{-1}\partial R/\partial T$ and the thermal uncertainty $\sigma_T(T)$ (resolution), calculated based on uncertainty propagation (see section S5 of the SI). $S_r(T)$ and $\sigma_T(T)$ are adequate parameters to compare different types of thermometers. Within the optimum window, the thermal resolution is $\pm$ 0.5 K and the relative sensitivity ranges between 25\%K$^{-1}$ and 10\%K$^{-1}$, typical values for temperatures in cryogenic ranges. For higher temperatures, $\sigma_T$ increases since it scales with $S_r^{-1}$. For lower temperatures, $\sigma_T$ also increases because the signal-to-noise ratio of the luminescence signals decrease.

\subsubsection{Self-referenced calibration using the LIR curve inflection point}

The function $R_\text{Stark}(T)$ [Eq. \eqref{eq:LIR_Stark}] exhibits an inflection point at $T_\text{infl} = \Delta E/2 k_\text{B}$, corresponding to the temperature where the absolute sensitivity $S_a = \partial R_\text{Stark} / \partial T$ reaches its maximum.  It is the upper bound of the optimum temperature measurement window $[\Delta E / (2+\sqrt{2})k_\text{B}, \; \Delta E / 2k_\text{B}]$. \cite{Suta_Meijerink_2020}

Similarly, the function $R_\text{Stark}(P)$ will also exhibit an inflection point at $P_\text{infl} = (T_\text{infl}-T_0)/a$. This occurs because $\partial^2 R/\partial P^2 = a^2 \cdot \partial^2 R/\partial T^2$. By acquiring a set of spectra for a range of $P$ values around $P_\text{infl}$, one can locate the inflection point $R_\text{Stark}(P_\text{infl})$. At this point, $T = T_\text{infl}$, allowing the inversion of Eq. \eqref{eq:LIR_Stark} to determine $C$ as $C = R_\text{Stark}(P_\text{infl}) \cdot \mathrm{e}^2$. This approach requires that the linear $T(P)$ relation remains valid within a suitable range around $P_\text{infl}$. Notably, knowledge of $a$ and $T_0$ is unnecessary, as the calibration parameter $C$ depends solely on identifying $R_\text{Stark}(P_\text{infl})$.

This identification can be done, for instance, by second-order differentiation of the experimental data with regard to $P$. A common technique is the Savitzky–Golay filtering \cite{Savitzky_1964}, which fits local subsets of data points with low-order polynomials using linear least squares and applies the result via convolution. The individual polynomials are then differentiated individually. Increasing the number of points improves smoothing but introduces bias, potentially shifting the apparent inflection point away from $P_\text{infl}$ \cite{Press_1990}.

Furthermore, since the FWHM of the $S_a$ curve scales linearly with $\Delta E$, the uncertainty in obtaining its maximum also increases proportionally. Therefore, the accuracy of thermometers operating optimally in higher temperatures (high $\Delta E$) will be lower.

\begin{figure}[!h] % SinSignal gle column figure
\begin{center}
	\includegraphics[width=\columnwidth]{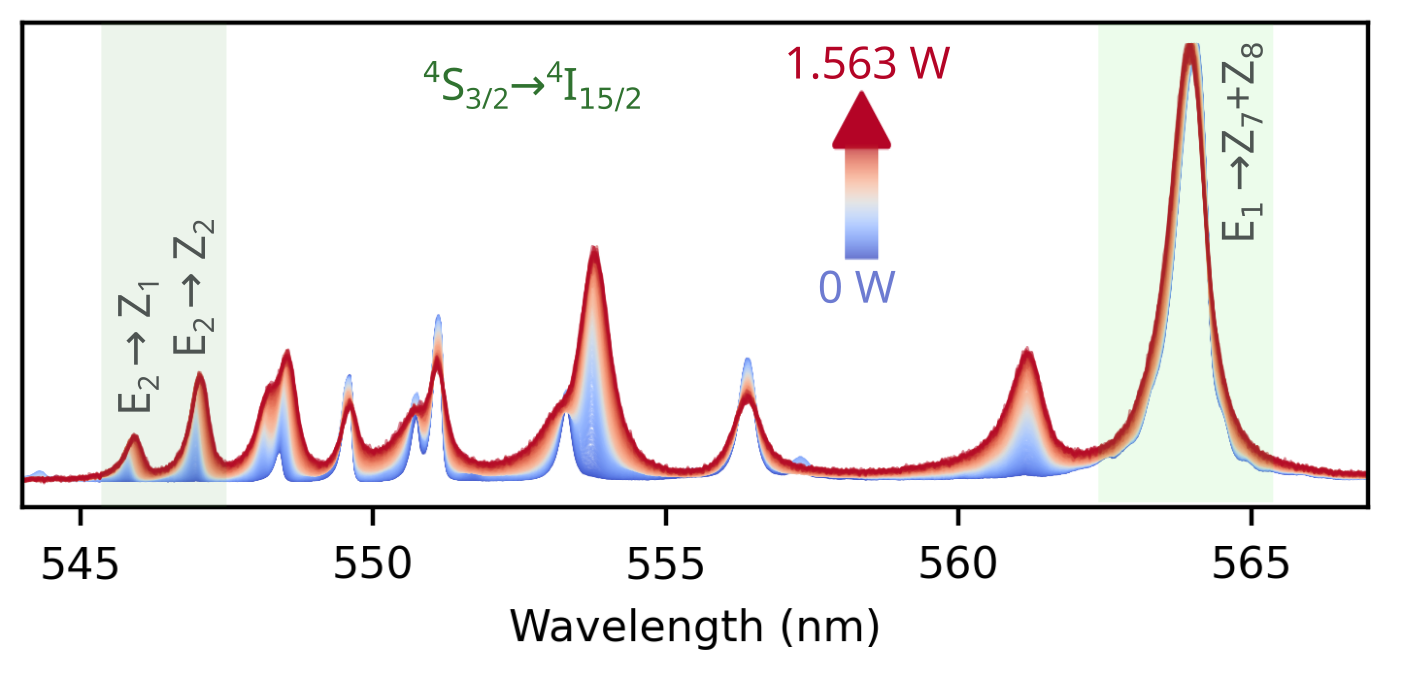}
	\caption{Sweep of the resulting luminescence spectra of the transition $^4$S$_{3/2}$ $\rightarrow$ $^4$I$_{15/2}$ as a function of the heat transfer rate, $P$. Counts are normalized to the maximum peak at 564 nm for visualization. The count rate of the maximum peak is 80 counts/s at room temperature. The shaded spectral areas highlight the lines used for absolute temperature measurements.}
	\label{fig:spectrum_Stark_green}
\end{center}
\end{figure}

\begin{figure*}[!ht] % Single column figure
\begin{center}
	\includegraphics[width=\textwidth]{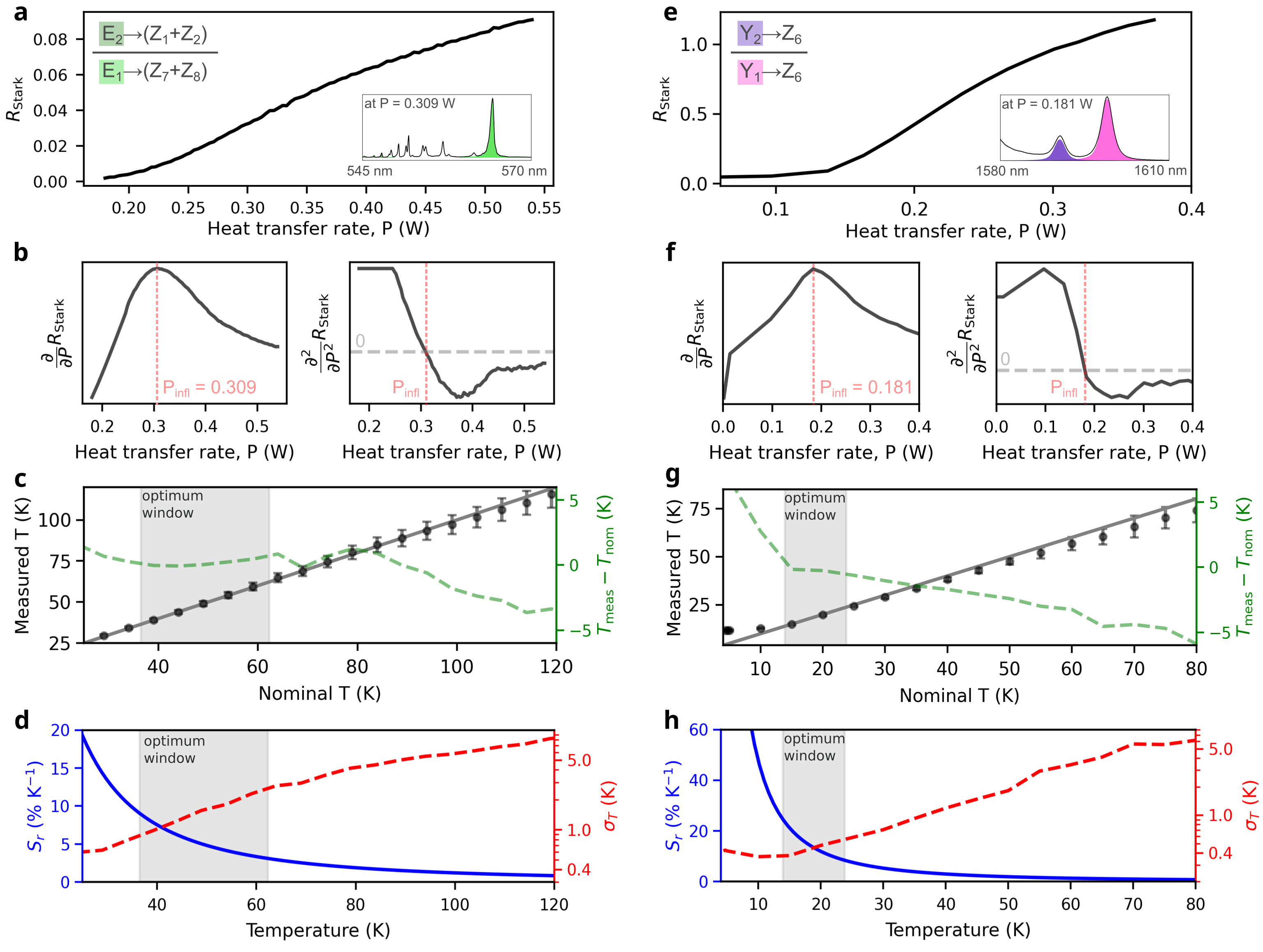}
	\caption{Self-referencing via the inflection point method. Left panel by using E$_1$ and E$_2$ Stark sublevels of $^4$S$_{3/2}$ and right panel by using Y$_1$ and Y$_2$ Stark sublevels of $^4$I$_{13/2}$. \textbf{a} and \textbf{e}: $R_\text{Stark}$ curves as a function of the external control parameter. Insets show the spectra at $P_\text{infl}$ with the Voigt fittings of the Stark lines used. \textbf{b} and \textbf{f}: show the first and second derivatives of the respective experimental curves obtained with a Savitzky-Golay algorithm. The horizontal dashed lines on the second derivative plots mark where it is zero. \textbf{c} and \textbf{g}: Comparison between temperature measurements through the luminescence spectra (measured T) with the internal cryostat references (nominal T). The solid straight lines represent the function $y=x$. The right-hand y-axes show the difference between these two quantities.  \textbf{d} and \textbf{h}: Characterization of the optical nanothermometers showing the relative sensitivities and thermal uncertainties.}
	\label{fig:inflection_cal}
\end{center}
\end{figure*}

We experimentally demonstrate the feasibility of finding the inflection point using the Savitzky-Golay differentiation for two independent pairs of TC Stark sublevels: i) E$_1$ and E$_2$ from the $^4$S$_{3/2}$ manifold, resulting in emission lines in the green spectral range \cite{Possmayer_2026}, and ii) Y$_1$ and Y$_2$ from the $^4$I$_{13/2}$ manifold resulting in emission lines in the NIR spectral range \cite{Boldyrev_2024}. The luminescence lines from E$_1$ and E$_2$ sublevels to the ground state are shown in Fig. \ref{fig:spectrum_Stark_green}, while those from Y$_1$ and Y$_2$ were already presented in Fig. \ref{fig:spectrum_Stark_IR}\textbf{b}. Section S4 of the SI shows the temperature-dependence characterization for the green spectral band. In particular, the energy difference between E$_1$ and E$_2$ is $\Delta E_{E_1E_2} = $ 86.6 $\pm$ 0.2 cm$^{-1}$ at 50 K, thus making the optimum temperature measurement window 36 -- 62 K\cite{Suta_Meijerink_2020}.

Fig. \ref{fig:inflection_cal}\textbf{a} shows $R_\text{Stark}(P)$ for E$_1$ and E$_2$ around $P_\text{infl}$, while Fig. \ref{fig:inflection_cal}\textbf{b} presents its first and second derivatives. The derivatives of the experimental curve were computed using a second-order Savitzky-Golay filter with 40 points per subset. A total of 100 spectra has been acquired around $P_\text{infl}$. The curve exhibits an inflection point at $P_\text{infl} = 0.309 \pm 0.001$ W. We then measure $R_\text{Stark}(P_\text{infl}) = 0.0353 \pm 0.002$  to obtain $C = 0.261 \pm 0.015$, in agreement with the value determined independently in Fig. S7\textbf{b} (See section S4 of the SI). By measuring $R_\text{Stark}$ and $\Delta E$ for a given spectrum, and considering $C = 0.261 \pm 0.015$, one has all necessary parameters to obtain $T$ from Eq. \eqref{eq:LIR_Stark}. Fig. \ref{fig:inflection_cal}\textbf{c} shows the measured temperatures, plotted against the nominal temperatures. Excellent agreement is observed within the optimum range, with temperature accuracy better than 0.5 K. In Fig. \ref{fig:inflection_cal}\textbf{d} we show the relative sensitivity and the temperature uncertainties calculated based on error propagation (Section S5 of the SI). The minimum uncertainty is 0.7 K, while relative sensitivities as high as 9.0 \% K$^{-1}$ have been obtained in the optimum window.

Figs. \ref{fig:inflection_cal}\textbf{e}-\textbf{h} present analogous results for $R_\text{Stark}(P)$ obtained from the Y$_1$ and Y$_2$ Stark sublevels of the $^4$I$_{13/2}$ manifold. We obtained 17 spectra around $P_\text{infl}$. The second-order Savitzky-Golay filter uses 6 points per subset. In this case, we obtain $P_\text{infl} = 0.181 \pm 0.001$ W, and $C = 2.27 \pm 0.06$, in agreement with that obtained with the previous method of calibration ($C = 2.3 \pm 0.1$), and also in agreement with that determined through temperature-dependent verification in Fig. S4\textbf{b}. Similarly, we observe accuracies better than 0.5 K in the optimum window and a minimum uncertainty of 0.4 K. The maximum relative sensitivity in the optimum window is 25 \% K$^{-1}$.

\section{Discussion}
We demonstrate robust absolute primary optical nanothermometric measurements using individual Stark sublevels in a rare-earth-doped solid. We developed two independent self-referenced calibration procedures that do not rely on external thermometric probes for the thermodynamic temperature determination: 1) exploiting the high-temperature limit of the $P \cdot \ln R_\text{Stark}(P)$ curve, enabling absolute thermometry optimally in the cryogenic regime; and 2) identifying the inflection point of the $R_\text{Stark}(T)$ curve, which occurs at $T_\text{infl} = \Delta E/2k_\text{B}$. To validate both approaches, we used Yb$^{3+}$,Er$^{3+}$-codoped yttrium oxide (Y$_2$O$_3$) nanoparticles, which exhibit well-resolved Stark spectral lines. The precise design of the host matrix is essential to ensure a sufficiently large Stark splitting and narrow spectral linewidths. In particular, the choice of yttria (Y$_2$O$_3$) as host matrix allows to observe Stark-Stark transitions with linewidths of 0.8 nm, enabling individual lines to be clearly spectrally resolved, identified and integrated \cite{Possmayer_2026, Galvao_2021}.

Two independent Stark sublevel pairs were investigated: E$_1$/E$_2$ from the $^4$S$_{3/2}$ manifold and Y$_1$/Y$_2$ from the $^4$I$_{13/2}$ manifold. The former pair produces green luminescence and enables absolute thermometry with an accuracy of 0.5 K within the optimal range of 36 -- 62 K, whereas the latter emits in the NIR and achieves 0.5 K accuracy in the 14 -- 24 K range. Despite the fact that green light is easier to detect, near-infrared light is more adequate to the study of biological samples. In any case, other Stark sublevel pairs can be selected provided that their spectral lines are spectrally resolvable, and that the thermalization dynamics follows Boltzmann statistics. This allows operation over different temperature ranges, including possibly near room temperature. 

%%%%%%%%%%%%%%%%%%%
A central assumption of this work is that Eq. \eqref{eq:LIR_Stark} remains valid over a suitable temperature window. This translates into conditions such as: 1) $\Delta E$ and $C$ are temperature-independent; 2) no additional photophysical processes effectively compete with electron–phonon thermalization; and 3) the excitation power is low enough to avoid non-negligible self-heating. Regarding condition 1): The radiative decay rates of Ln$^{3+}$ systems are known to exhibit slight temperature dependence, but in most cases this variation is negligible \cite{Suta_Meijerink_2020, Ciric_2020}, allowing $C$ to be approximated as temperature-invariant. Likewise, matrix conformational changes with temperature can modify the Stark splitting structure, thus altering $\Delta E$. In the SI material, we presented $\Delta E$ vs. $T$ measurements showing variations on the order of $3\times 10^{-3}$ cm$^{-1}$/K; as a result, $\Delta E$ can also be considered as temperature-invariant depending on the targeted accuracy of the absolute thermometer. Concerning conditions 2) and 3): Competing photophysical processes such as cross-relaxation or excited-state absorption would modify the LIR \textit{vs.} $T$ curve, deviating it from the simple exponential dependence in Eq. \eqref{eq:LIR_Stark} \cite{Pessoa_2025}. Similarly, self-heating -- more pronounced at lower temperatures for a fixed laser power -- would also distort $R_\text{Stark}$ from Eq. \eqref{eq:LIR_Stark}. To avoid these artifacts, we carefully controlled the synthesis route, doping concentration and size of the nanoparticles during chemical assembly \cite{Possmayer_2026}, as well as the excitation power densities during experiments. The controlled temperature-dependent measurements of $R_\text{Stark}$, presented in sections S3 and S4 of the SI, assure that $R_\text{Stark}$ follows a plain Boltzmann distribution over a wide temperature range, with the correct $C$ and $\Delta E$ parameters.

Single-particle measurements have been applied for relative primary nanothermometry \cite{Goncalves_2021, Pessoa_2023, Galvao_2021}, but it has been shown that different particles can have slight different calibration parameters, even stemming from the same batch \cite{Goncalves_2021, Galindo_2021, Galindo_2021_corr}. Therefore, using the methods developed in this work for single nanoparticles would allow, for instance, determining the local thermodynamic temperature in the nano- or microscale without disturbing its behavior with an external probe, specially in environments with multiple heat sources or sinks, as, for instance in \textit{in vivo} biological samples. For example, a contactless absolute nanothermometer can be used to eliminate several artifacts, possibly helping to close the $10^{5}$ gap in single cell thermometry \cite{Sotoma_2021, Suzuki_2020}. Furthermore, because the thermalization of rare-earth ions with their local environment occurs at the single-ion level, single-ion thermometry could provide a route toward absolute temperature measurements using quantum emitters. Precision can be further enhanced when single-qubit thermometers exploit quantum coherence and entanglement \cite{Jevtic_2015}. Such capabilities open new opportunities for quantum thermometry and quantum thermodynamics investigations, for example in probing heat transport and dissipation in nanoscale sub-kelvin electronic circuits \cite{Pekola_2015, Gasparinetti_2015}.

It is worth mentioning that the approach developed in this work is not suitable if one uses the full manifold-to-manifold intensity band to calculate the LIR ($R_\text{manifold}$) because the energy difference governing Boltzmann equation [Eq. \eqref{eq:LIR_Stark}] should be treated as an effective value, which depends not only on the energy-level structure of the Ln$^{3+}$ ion, but also on all radiative decay rates of the lines comprising the luminescence band \cite{Pessoa_2025}. This effective energy difference is different from the barycenter energy difference \cite{Possmayer_2026}, meaning that it is not possible to extract it solely from the luminescence spectrum without a previous temperature calibration.

\section{Materials and Methods}
\subsection{Experimental details}

We employed a powder sample of Y$_2$O$_3$: Yb$^{3+}$ (1.5 mol\%) / Er$^{3+}$ (0.5 mol\%) nanoparticles with an average diameter of 80 nm. Sample's characterization details as morphology and X-ray diffractogram can be found in ref. [\!\!\citenum{Possmayer_2026}].

The dry powder was pressed into a copper sample holder and mounted inside a closed-cycle cryostat (Cryostation\textsuperscript{\textregistered} s50, Montana Instruments). An internal temperature sensor (Cernox\textsuperscript{\textregistered} CX-1070, Lake Shore Cryotronics) serve as a reference for comparison. The manufacturer calibration curve displays a typical accuracy of $\pm$6 mK at 10 K and $\pm$48 mK at 300 K \cite{LakeShore_AppendixD_2024}. By tuning the heat transfer rate (input power), the internal cryostat temperature can be changed from 4.4 K to 350.0 K, according to the internal probe.

The sample was excited by a femtosecond laser source emitting at 980 nm with 80 MHz repetition rate and 7 nm of spectral pulse width. The light excitation irradiance is 2.0 kW/cm$^2$. The Yb$^{3+}$ ions work as sensitizers and the observe luminescence spectra stem from electronic transitions of the Er$^{3+}$ ions \cite{Goncalves_2021}. We performed measurements for both green and NIR spectral regions. The green luminescence spectrum originates from an upconversion process via a well-known Yb$^{3+}$ $\rightarrow$ Er$^{3+}$ two-photon energy transfer mechanism \cite{Goncalves_2021}, whereas the NIR emissions arises primarily from non-radiative relaxations from $^4$I$_{11/2}$ to $^4$I$_{13/2}$ after a single-photon energy transfer process \cite{Boldyrev_2024}. Emission was collected in reflection geometry using a lens placed inside the cryostat. The resulting emitted light was directed to a spectrometer with a high-resolution grating (1800 grooves/mm) and detected by a cooled CCD camera. Integration time for the green and NIR spectrum was 60 s and 20 s, respectively. Further experimental details, including a diagram of the experimental setup, can be found in a previous work from our group, ref. [\!\!\citenum{Possmayer_2026}]. 

\bmhead{Data availability} The data that support the findings of this study are available from the corresponding author upon reasonable request.

\backmatter

\bmhead{Supplementary information}

The temperature dependence of the material’s luminescence spectra in the green and NIR spectral ranges is presented in the Supplementary Information. A detailed analysis of the linearity between temperature and the heating device’s power input is also provided, together with a discussion of accuracy versus precision in the context of the methods employed in this work.

\bmhead{Acknowledgements}
A. R. Pessoa, J. A. O. Galindo and A. M. Amaral acknowledge FINEP (Call MCTI/FINEP/FNDCT/CENTROS TEMÁTICOS 2023 grant number 1020/24). L. F. dos Santos acknowledges FAPESP (grant numbers: 2020/04157-5 and 2023/03092-5). R. R. Gonçalves acknowledges CNPq (grant numbers: 303110/2019-8 and 306191/2023-7) and FAPESP (grant number: 2021/08111-2). L. de S. Menezes and T. Possmayer acknowledge the support from the Center for Nanoscience (CeNS), Ludwig Maximilians-Universit\"at M\"unchen, Germany, and the Bavarian program Enabling Quantum Communication and Imaging Applications (EQAP). We acknowledge Brazilian National Institutes of Science and Technology (INCTs), INCT-INFO (National Institute of Photonics) and INCT LumiNanoTec Tecnologias Luminescentes: Sensores, marcadores e nanodispositivos integrados.

\bmhead{CRediT author statement} A. R. PESSOA: Conceptualization, Formal analysis, Writing- Original draft; T. POSSMAYER: Investigation, Formal analysis, Writing- Original draft; J. A. O. GALINDO: Formal analysis, Writing- review \& editing; L. F. dos SANTOS: Resources, Writing- review \& editing; R. R. GONÇALVES: Resources, Writing- review \& editing; L. de S. MENEZES: Supervision, Project Administration, Writing- review \& editing; A. M. AMARAL: Conceptualization, Project Administration, Writing- review \& editing.

\bmhead{Conflict of interest statement} The authors declare no conflict of interest.

\pagebreak

\bibliography{sn-bibliography}% common bib file
%\onecolumn{
%\bibliography{references}}



\section*{Supplementary material}

# S1 Relation between the temperature of the particle and the heat transfer rate

The thermal interface between a heating/cooling element and a nanoparticle thermometer, as depicted in Fig. S1a, can be modeled by a thermal circuit shown in Fig. S1b. The heat source delivers heat at the rate $\dot{Q}$. Fourier's law of heat conduction leads to an analogy to Ohm's law for electrical circuits[1]: $\Delta T = R \cdot \dot{Q}$, where $\Delta T$ is the temperature difference between two points in space. $\Delta T$ is the analogous of voltage while $\dot{Q}$ is the analogous of electrical current. $R$ is the thermal resistance of the material, or at interfaces.

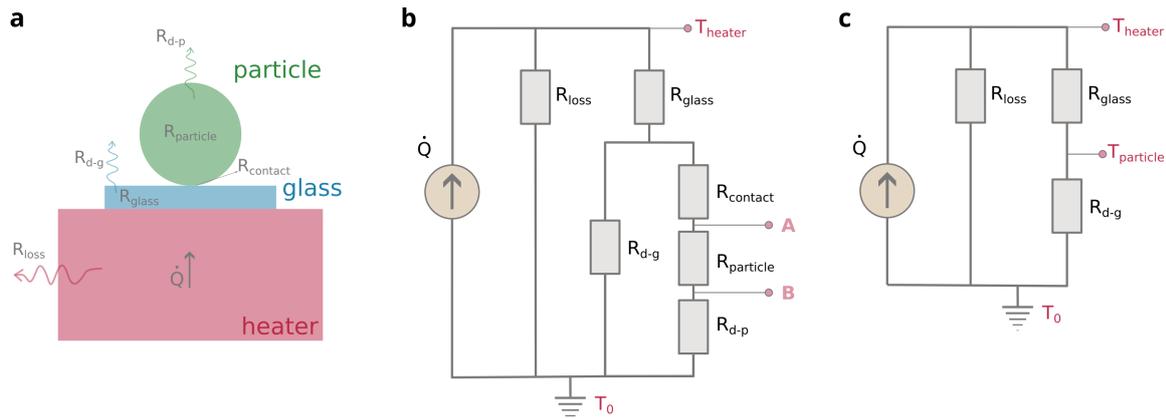

**Fig. S1** **a**: Schematics of the interface between a heater element and a nanoparticle thermometer. **b**: Thermal circuit modeling the situation presented in **a**. **c**: Simplification of the thermal circuit presented in **b** to allow easier physical interpretation.

The heater dissipates power to its surroundings with resistance $R_{\text{loss}}$ and to the glass + particle system with equivalent resistance $R_{\text{glass}} + R_{\text{d-g}}(R_{\text{contact}} + R_{\text{particle}} + R_{\text{d-p}})/(R_{\text{d-g}} + R_{\text{contact}} + R_{\text{particle}} + R_{\text{d-p}})$. Here, $R_{\text{glass}}$ is the thermal resistance of the glass, $R_{\text{d-g}}$ is the thermal resistance representing the heat dissipation by the glass through its surface, $R_{\text{contact}}$ is the thermal contact resistance between the particle's and the glass, $R_{\text{particle}}$ is the thermal resistance of the nanoparticle, and $R_{\text{d-p}}$ is the thermal resistance representing the heat dissipation by the particle.

Due to the nanoscale size of the particle, $R_{\text{particle}} \ll R_{\text{d-p}}$ (small Biot numbers[1]), meaning that the temperature inside the particle can be considered uniform. So, either point A or B in Fig. S1b can represent the particle's temperature ($R_{\text{particle}} \to 0$). Furthermore, the heat dissipated through the glass is much greater than the heat dissipated through the particle, meaning that $R_{\text{d-g}} \ll R_{\text{d-p}}$ and $R_{\text{d-g}} \ll R_{\text{contact}}$. As a result, we can simplify the thermal circuit to that presented in Fig. S1c, only as means for obtaining easier physical understanding. In this case, $T_{\text{ptc}}$ is written as

$$T_{\text{ptc}} = T_0 + \frac{R_{\text{d-g}} \cdot R_{\text{loss}}}{R_{\text{glass}} + R_{\text{d-g}} + R_{\text{loss}}} \dot{Q}, \tag{S1}$$

thus showing that $T = aP + T_0$ should be a valid prescription to describe our experimental system. $T_0$ can be the room temperature, for instance, or the helium boiling point temperature if the system is placed inside a cryostat, as it is the case of the present study. It is worth noticing that any configuration of heat or temperature sources with linear thermal resistors would lead to the same linear result, $T = aP + T_0$, since it is a fundamental consequence of the the heat equation $\nabla^2 T = 0$ in steady state and homogeneous materials, provided that, for suitable temperature ranges, we consider that the thermal resistances are temperature-independent.

Even if the thermal resistances change with the temperature, the linear relation should still be verified for a suitable range around a given temperature point $T_c$. Using the Norton equivalent circuit for Fig. S1c, we can write $T_{\text{particle}} = T_0 + R_{eq}(T_{\text{particle}})\dot{Q}$. For small temperature variations around $T_c$, $T_{\text{particle}} = T_c + \delta T$, $\dot{Q} = \dot{Q}_c + \delta\dot{Q}$ and $R_{eq} = R_c + R'\delta T$, such that

$$\delta T = \frac{R_c \delta \dot{Q}}{1 - R'\dot{Q}} \quad . \tag{S2}$$

For $\dot{Q}_c \gg \delta\dot{Q}$,

$$\delta T = \frac{R_c \delta \dot{Q}}{1 - R'\dot{Q}_c} \quad . \tag{S3}$$

Thus, the overall relation $T = aP + T_0'$ should be valid in a small neighborhood close to the calibration temperature $T_c$ as long as there are no abrupt changes in the thermodynamics properties of the thermal circuit elements (such as a phase transition). Moreover, notice that the offset temperature $T_0' \neq T_0$ in general due to the temperature dependence of the thermal resistance, $R'$.

## S2 Linear relation between $P$ and the cryostat internal temperature

The cryostat allows controlling the input power of its internal heaters to increase the temperature above the helium boiling point. In the main text, we assumed that the relation between the input power and the internal temperature is linear for a suitable temperature range. Figure S2 shows this relation for the complete system's working range.

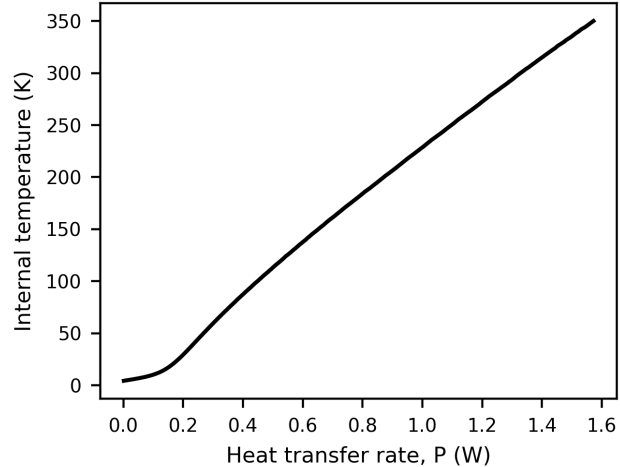

**Fig. S2** Measured relation between cryocooler input power (energy transfer rate) and its internal temperature.

## S3 Characterization of the Stark lines $Y_1 \rightarrow Z_6$ and $Y_2 \rightarrow Z_6$

The Stark sublevels $Y_1$ and $Y_2$ belong to the $^4I_{13/2}$ manifold of the $Er^{3+}$ ions, as depicted in Fig. 1a of the main manuscript. They are used as pair of thermally coupled sublevels for absolute primary thermometry measurements by monitoring the lines $Y_1 \rightarrow Z_6$ and $Y_2 \rightarrow Z_6$, where $Z_6$ is a given Stark sublevel of the ground state manifold ($^4I_{15/2}$).

The selected Stark lines were deconvolved from the luminescence spectrum using a Voigt profile fitting, according to the procedures detailed in ref. [2]. The partially superposed emission lines were fitted with the function

$$V(x; V_0, x_0, \sigma, \gamma) = V_0 \int_{-\infty}^{\infty} G(x'; \sigma) L(x - x'; \gamma) \, dx', \tag{S4}$$

where $G(x; x_0, \sigma) = \frac{1}{\sigma\sqrt{2\pi}} \exp\left(-\frac{(x-x_0)^2}{2\sigma^2}\right)$ is the Gaussian component and $L(x; x_0, \gamma) = \frac{\gamma/\pi}{(x-x_0)^2+\gamma^2}$ is the Lorentzian component. $V_0$, $x_0$, $\sigma$ and $\gamma$ are fitting parameters.

Fig. S3a shows a luminescence spectrum in the wavelength range of interest along with the individual contributing Voigt profiles. The sum of all Voigt peaks composes the fitted spectrum. The residue of the fitted spectrum is shown in Fig. S3b and they are smaller than 4 % in the whole spectral range considered. Since the transitions $Y_1 \to Z_6$ and $Y_2 \to Z_6$ share the same final state ($Z_6$), the difference between their energy barycenters is equal to the 'real' energy difference between $Y_1$ and $Y_2$ states ($\Delta E_{\text{bary}} = \Delta E_{Y_1 Y_2}$). To calculate $\Delta E_{\text{bary}}$ one needs to transform the spectral lines from the wavelength representation $I(\lambda)$ to the energy representation $\tilde{I}(E)$, then obtain the individual barycenters by $E_{\text{bary}} = \left(\int_{-\infty}^{\infty} E \cdot \tilde{I}(E) dE\right) / \left(\int_{-\infty}^{\infty} \tilde{I}(E) dE\right)$, as described in ref. [3]. Fig. S3c shows the measured $\Delta E_{Y_1 Y_2}$ according to this procedure as a function of the internal cryostat temperature, for characterization. As the temperature increases, the host lattice expands changing the effective crystal field on the $Er^{3+}$ ions, hence shifting the energy of the Stark sublevels.

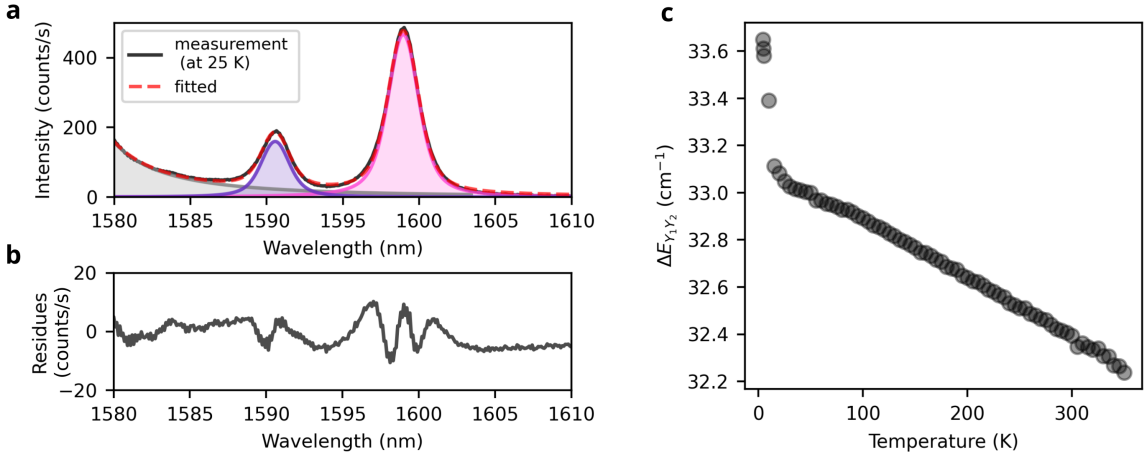

**Fig. S3** **a**: Partial luminescence spectrum of the transition $^4I_{13/2} \to {}^4I_{15/2}$ in $Y_2O_3$: $Yb^{3+}/Er^{3+}$ systems showing the $Y_1 \to Z_6$ and $Y_2 \to Z_6$ Stark lines and their respective fitted Voigt curves. **b**: Fit residue from the experimental curves. **c**: Energy difference between $Y_1$ and $Y_2$ states as a function of the temperature.

Keeping the excitation power at reasonable values is essential to avoid effects such as self-heating, excited state absorption, and stimulated emission, which could jeopardize accurate absolute temperature measurements. We measured the luminescence intensity of the corresponding $Y_1 \to Z_6$ and $Y_2 \to Z_6$ lines as a function of the excitation power at room temperature (Fig. S4a). We observe that the integrated intensities $I$ follow a power-law $I \propto P^n$, where $P$ is the excitation irradiance (excitation power divided by the excitation area). As expected for the $^4I_{13/2} \to {}^4I_{15/2}$ transition below the saturation regime, we obtained $n = 1$.

To experimentally verify Boltzmann assumption (Eq. (1) of the main text), Fig. S4b shows $R_{\text{Stark}}$ as a function of the internal cryostat temperature, measured with its internal thermometric probe. The calculated $\Delta E_{\text{eff}}$ from the fitting[3] agrees with $\Delta E_{Y_1 Y_2}$ for temperatures above 15 K. For temperatures below this point, we observe a plateau in the $\ln(R_{\text{Stark}})$ vs. $T^{-1}$ curve, where Boltzmann distribution is not attained. This is most probably due to the competition between radiative and non-radiative decay rates, as discussed in ref. [3]. Therefore, a Boltzmann thermometer using this pair of levels would not be appropriate for $T < 15K$.

S4

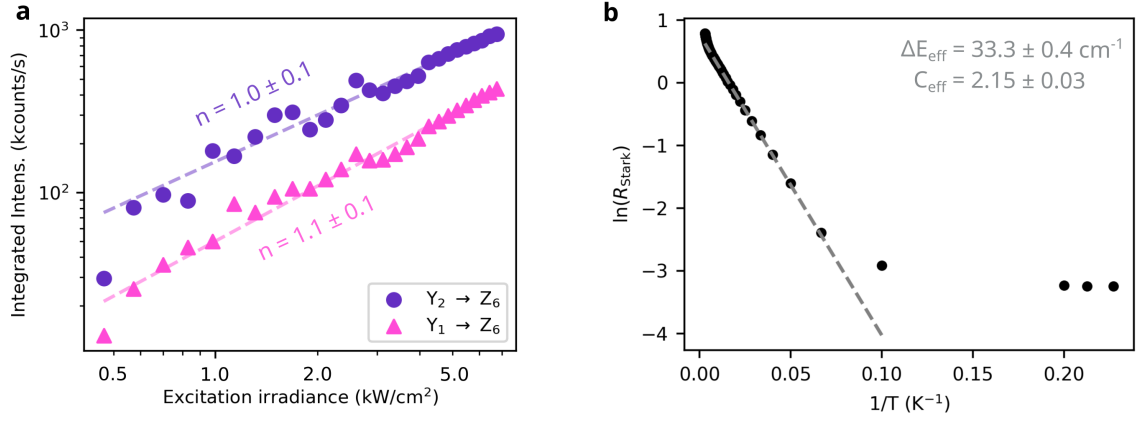

**Fig. S4** **a**: Intensity of the Stark lines as a function of the excitation irradiance. The dashed curves represent the fittings $I = A \cdot P^n$, with $A$ and $n$ as fitting parameters. The resulting $n$ values are shown within the plot. **b**: Experimental verification of Boltzmann assumption. The dashed line represents a linear fitting for tempertures higher than 15 K, from which we extract $C_{\text{eff}}$ and $\Delta E_{\text{eff}}$ (inset).

## S4  Characterization of the Stark lines $E_1 \rightarrow (Z_7 + Z_8)$ and $E_2 \rightarrow (Z_1 + Z_2)$

Two-photon upconversion processes with excitation at 980 nm can populate the $^2H_{11/2}$ and $^4S_{3/2}$ spin-orbit manifolds of the Er$^{3+}$ ions (Fig. S5a). These manifolds decay radiatively to the ground state ($^4I_{15/2}$), resulting in luminescence bands in the green spectral region. Fig.S5b shows the transitions $^2H_{11/2} \rightarrow {}^4I_{15/2}$ and $^4S_{3/2} \rightarrow {}^4I_{15/2}$ for temperatures between 4.4 and 350.0 K. As discussed in the main text, we have selected the Stark lines $E_1 \rightarrow (Z_7 + Z_8)$ and $E_2 \rightarrow (Z_1 + Z_2)$ of the $^4S_{3/2}$ band (highlighted in Fig. S5c) as thermally coupled pairs for performing absolute primary thermometry.

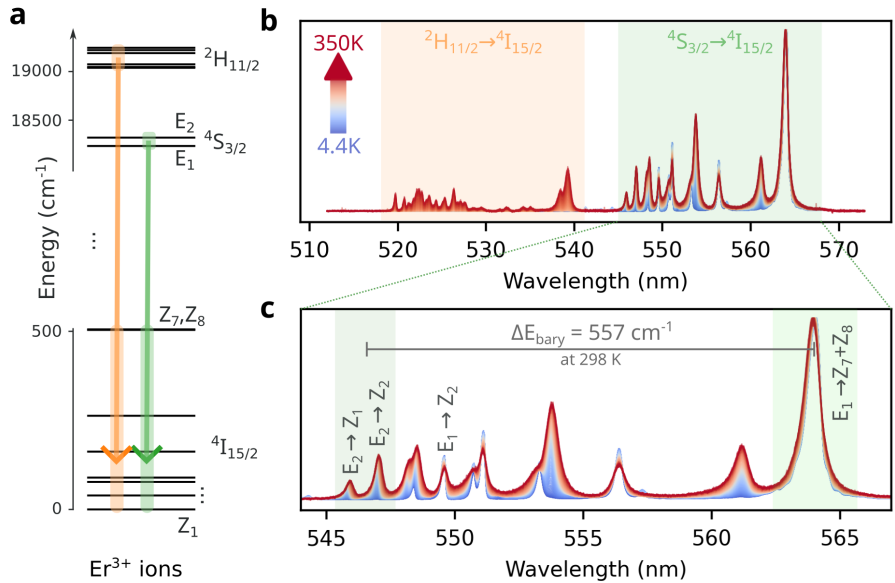

**Fig. S5**  **a**: Simplified energy-level diagram of Er$^{3+}$ ions showing the $^2H_{11/2}$ and $^4S_{3/2}$ manifolds, which are electronically populated after a two-photon excitation at 980 nm. **b**: Luminescence spectra of the transitions $^2H_{11/2} \rightarrow {}^4I_{15/2}$ and $^4S_{3/2} \rightarrow {}^4I_{15/2}$ as a function of the internal temperature of the cryostat. **c**: The same plot as in **b**, but highlighting the Stark lines that will be used for absolute primary optical thermometry.

The selected Stark lines are depicted in Fig. S6a and were fitted similarly as in Section S1. The fit residues are shown in Fig. S6b. Since $E_1 \rightarrow (Z_7 + Z_8)$ and $E_2 \rightarrow (Z_1 + Z_2)$ lines do not share

the same final state, $\Delta E_{\text{bary}}$ is different from $\Delta E_{E_1 E_2}$. To properly measure $\Delta E_{E_1 E_2}$ we chose the pair of Stark lines $E_1 \rightarrow Z_2$ and $E_2 \rightarrow Z_2$, as shown in Fig S5c. $\Delta E_{E_1 E_2}$ is depicted in Fig. S6c for temperatures between 4.4 K and 350.0 K, for characterization.

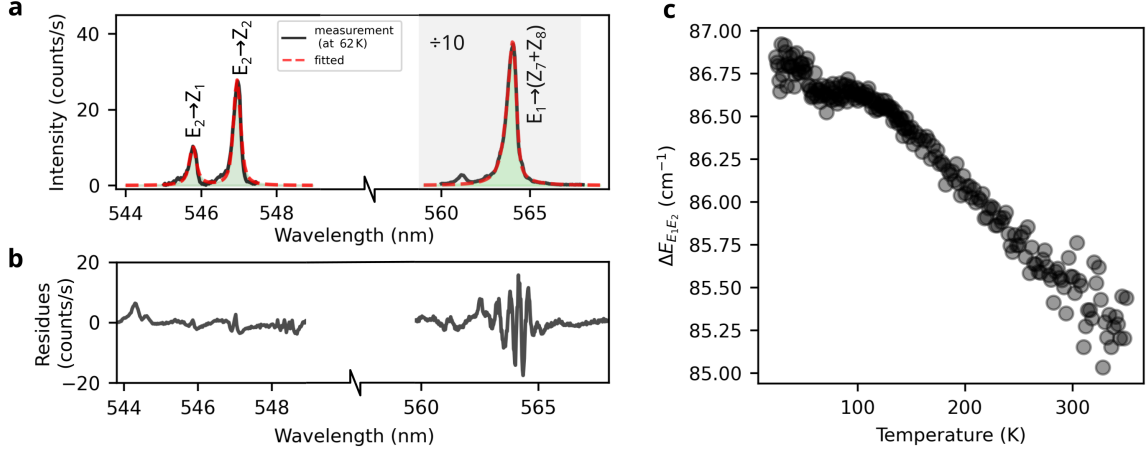

**Fig. S6** **a**: Partial luminescence spectrum of the transition $^4S_{3/2} \rightarrow {}^4I_{15/2}$ in $Y_2O_3$: $Yb^{3+}/Er^{3+}$ systems showing the $E_1 \rightarrow (Z_7 + Z_8)$ and $E_2 \rightarrow (Z_1 + Z_2)$ Stark lines and their respective fitted Voigt curves (see main text for details). **b**: Fit residues from the experimental curves. **c**: Energy difference between $E_1$ and $E_2$ as a function of the temperature.

Power-law analysis, similarly to that of section S1, reveals $n = 2$ for both transitions (Fig. S7a), indicating a non-saturated upconverting process. Similarly, the plot $\ln(R_{\text{Stark}})$ vs. $T^{-1}$ curve (Fig. S7b), which serve as a verification of the Boltzmann assumption, shows $\Delta E_{\text{eff}} = 83.8 \pm 0.4$ cm$^{-1}$, in excellent agreement with $\Delta E_{E_1 E_2}$ spectrocopically measured.

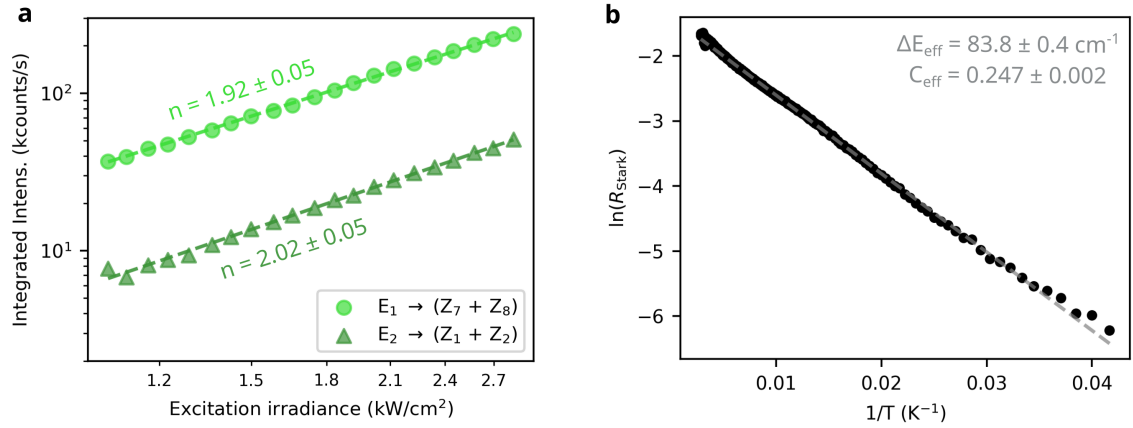

**Fig. S7** **a**: Intensity of the Stark lines as a function of the excitation irradiance. The dashed lines represent the fittings $I = A \cdot P^n$, with $A$ and $n$ as fitting parameters. **b**: Experimental verification of Boltzmann assumption. The dashed line represents a linear fitting, from which we extract $C_{\text{eff}}$ and $\Delta E_{\text{eff}}$ (inset).

## S5 Accuracy and precision

In both proposed methods, the temperature uncertainty is obtained through standard error propagation from the uncertainties of $\Delta E$ and $C$. As discussed in the main text, $\Delta E$ is measured directly from the luminescence spectrum, with uncertainties on the order of 0.5 cm$^{-1}$ or less (depending on the resolution of the spectrometer), while $C$ depends on the specific method employed. It is important to note that the thermal uncertainty reflects the thermometer's precision, whereas accuracy

corresponds to the deviation between the measured and the actual temperature. Obtaining accuracy requires comparison of the proposed thermometric system with previously calibrated probes.

For absolute primary thermometry based on the measurement of the slope $\mathrm{d}(P \cdot \ln R_{\mathrm{Stark}})/\mathrm{d}P$ in the high-temperature limit (Eq. (3) of the main manuscript), the uncertainty in $\ln C$ is equal to the uncertainty resulting from the linear fitting to obtain the slope. Naturally, this uncertainty calculation should inherit the individual uncertainties of the data point $\sigma_{dp}$, as $\sigma_{dp}^2 = (P \cdot \frac{1}{R_{\mathrm{Stark}}}\sigma_R)^2 + (R_{\mathrm{Stark}} \cdot \sigma_P)^2$, where $\sigma_R$ is the uncertainty of $R_{\mathrm{Stark}}$ and $\sigma_P$ is the uncertainty of $P$. The errors on the power measurements, $\sigma_P$, depend on the current and voltage sensors used to obtain $P$. In our case we considered as $\sigma_P = 0.001$ W as given by the equipment.

To compute $\sigma_R$, a standard approach is to record multiple spectra at a fixed point and measure the mean and standard deviation of $R_{\mathrm{Stark}}$. Alternatively, as adopted in this work, the uncertainty can be estimated from the Voigt profile fittings of the spectrum. Given that $\int_{-\infty}^{\infty} V(x), \mathrm{d}x = V_0$, the uncertainty in the Stark line areas can be taken as the uncertainty of the fitted parameter $V_0$, as provided by the fitting routine. This estimate inherently accounts for the residuals of the experimental data and potential correlations among fitting parameters. It should be noted that smaller $V_0$ values are associated with larger relative uncertainties due to the shot-noise limit.

For absolute thermometry based on the measurement of $R_{\mathrm{Stark}}$ at the inflection point, the uncertainty in $C$ was considered equal to that of $R_{\mathrm{Stark}}(P_{\mathrm{infl}})$ multiplied by $\mathrm{e}^2$, i.e., $\sigma_C = \sigma_R \cdot \mathrm{e}^2$ (see main text). The specific error treatment on $\sigma_R$ depends on the method used to identify the inflection point. When using a Savitzky–Golay filter, the uncertainty depends on the local fitting residuals, and naturally the experimental $\sigma_R$ must be incorporated in the local fit. Then, $P_{\mathrm{infl}}$ is identified with some uncertainty, and the corresponding uncertainty in $R_{\mathrm{Stark}}(P_{\mathrm{infl}})$ is calculated through standard error propagation.

Regardless of the method used to obtain $\Delta E$ and $C$, the temperature measurement uncertainty is given by[3]

$$\sigma_T^2(T) = \frac{1}{S_r(T)^2} \left[ \left(\frac{\sigma_{\Delta E}}{k_{\mathrm{B}}T}\right)^2 + \left(\frac{\sigma_C}{C}\right)^2 + \left(\frac{\sigma_R}{R}\right)^2 \right]. \tag{S5}$$

where we disregarded the correlation between the variables because $\Delta E$ and $C$ are measured independently in the absolute primary thermometry approach.



\end{document}